\documentclass[]{aa}
\usepackage{psfig}
\def\apj{ApJ}

\def\ueber#1#2{{\setbox0=\hbox{$#1$}%
  \setbox1=\hbox to\wd0{\hss$ #2$\hss}%
  \offinterlineskip
  \vbox{\box1\box0}}{}}

\def\lesssim{\,\lower  1mm  \hbox{\ueber{\sim}{<}}\,}
\def\grsim{\,\lower  1mm  \hbox{\ueber{\sim}{>}}\,}

\begin{document}

\title{The frequency of occurrence of novae hosting an ONe white dwarf}

\author{Pilar Gil--Pons\inst{1}  \and  
        Enrique Garc\'{\i}a--Berro\inst{1,2} \and 
        Jordi Jos\'e\inst{2,3} \and 
        Margarita Hernanz\inst{2,4} \and
        James W. Truran\inst{5}}

\titlerunning{Frequency of ONe novae}  
\authorrunning{P. Gil--Pons et al.}

\institute{   
$^1$Departament de  F\'\i sica Aplicada,  Universitat Polit\`ecnica de
    Catalunya,  c/Jordi Girona  s/n, M\`odul  B-4, Campus  Nord, 08034
    Barcelona, Spain\\
$^2$Institute for  Space Studies  of Catalonia, c/Gran  Capit\`a 2--4,
    Edif.   Nexus  104, 08034  Barcelona,  Spain\\ 
$^3$Departament  de  F\'{\i}sica  i  Enginyeria  Nuclear,  Universitat
    Polit\`ecnica de Catalunya, Av.   V\'\i ctor Balaguer, s/n, 08800,
    Vilanova i la Geltr\'u (Barcelona), Spain\\
$^4$Institut de Ci\`encies de l'Espai (CSIC)\\
$^5$Department  of Astronomy  and  Astrophysics, 5640  S.  Ellis  Ave,
Chicago, IL~60637, USA}

\date{\today}

\abstract{In this  paper, we revisit the problem  of the determination
of  the  frequency of  occurrence  of  galactic  nova outbursts  which
involve  an  oxygen--neon (ONe)  white  dwarf.   The improvement  with
respect to previous work on the  subject derives from the fact that we
use  the results that  our evolutionary  calculations provide  for the
final mass  and for  the chemical profiles  of intermediate-to-massive
primary components of close  binary systems.  In particular, the final
evolutionary  stages, such  as  the carbon  burning  phase, have  been
carefully followed  for the  whole range of  masses of  interest.  The
chemical profiles obtained with  our evolutionary code are of interest
in  determining the  chemical composition  of the  ejecta  after being
processed  through  the  thermonuclear  runaway, although  such  other
factors as the efficiency of  the mixing between the accreted material
and that  of the underlying white  dwarf must also  be considered.  In
our  calculations of  the frequency  of occurrence  of  nova outbursts
involving  an  ONe  white  dwarf,   we  also  take  into  account  the
observational selection effects introduced by the different recurrence
times of the  outbursts and by the spatial  distribution of novae.  In
spite  of the  very  different evolutionary  sequences,  we find  that
approximately 1/3 of the novae  observed in outburst should involve an
oxygen-neon  white  dwarf,  in  agreement  with  previous  theoretical
estimates.
 }

\maketitle


\section{Introduction}

The  formation of  massive  oxygen--neon white  dwarfs in  cataclysmic
binaries and  the occurrence of ONe  novae was first  suggested -- and
its consequences investigated -- by  Law \& Ritter (1984), even before
any observational  evidence for such objects  was found. Subsequently,
the determination of  the frequency of occurrence of  ONe white dwarfs
in galactic classical  nova systems has also been  considered by other
authors  --- see,  for instance,  Truran  \& Livio  (1986), Ritter  et
al. (1991), and Livio (1992).   If the evolution of primary components
in  close binary  systems (CBSs)  were  identical to  those of  single
stars, and  if observational selection effects  were negligible (which
is not the case), one would  expect a peak in the mass distribution of
white dwarfs around the canonical value of $0.6\, M_ \odot$ (Bergeron,
Saffer \& Liebert 1992) found  for single white dwarfs.  We would thus
expect  to find a  small number  of oxygen-neon  white dwarfs  in nova
systems.  In  contrast, there is  an increasing body  of observational
information  indicating that  the  actual mass  distribution of  white
dwarfs in  galactic nova systems  is skewed toward higher  masses.  In
fact, the mass distribution for galactic nova systems is peaked around
$1.0\, M_\odot$ (Webbink, 1990).  Moreover, the observed abundances in
the ejecta  of some novae --- see,  e.g.  Gehrz et al.   (1998), for a
recent review  of the abundances  inferred from observations  of novae
--- show high neon enhancements (Saizar  et al., 1992; Andre\"a et al.
1994; Austin  et al., 1996;  Vanlandingham et al., 1997).   This again
strongly suggests, when compared  with theoretical models (Jos\'{e} \&
Hernanz, 1998), that the underlying  white dwarf is made of oxygen and
neon, and therefore that it is  a massive white dwarf.  We should thus
expect that a  significant fraction of nova systems  contain ONe white
dwarfs.

Ritter et  al.  (1991) estimated  that the frequency of  occurrence of
nova outbursts  hosting an ONe white  dwarf in the  Galaxy was between
$25  \%$ and  $57 \%$,  in good  agreement with  the previous  work by
Truran   \&  Livio   (1986),   and  compatible   with  the   available
observational data.  However, in a subsequent discussion of this issue
in  which the  observational uncertainties  were  further scrutinized,
Livio  \&  Truran (1994)  adopted  a  more  cautious approach  to  the
identification of ONe novae.  They  chose to classify novae into three
distinct groups. The first group corresponds to novae with high helium
abundances  and  modest  CNO  nuclei  enrichment.   The  second  group
includes  novae with  high  CNO nuclei  enrichment  and moderate  neon
enhancement.  Finally, the third group corresponds to those novae with
high  helium to  hydrogen  abundance  ratio and  both  neon and  heavy
element enrichment.  The well established ONe novae should then be the
ones  belonging to  the  third  group ---  constituting  two or  three
objects out of 18 according to Livio \& Truran (1994) --- whereas less
restrictive ONe nova identifications predict that ONe novae constitute
about one third  of the sample.  Whatever the  appropriate fraction of
ONe  novae  might   be,  it  is  worth  noting   at  this  point  that
observational selection  effects also play  a key role  in determining
this frequency (Livio \& Soker 1984; Ritter \& Burkert 1986; Ritter et
al., 1991).

Another  factor  that  has   a  considerable  influence  on  the  mass
distribution of novae is the evolutionary history of the binary system
previous  to the  formation  of the  nova.   Indeed, the  evolutionary
history of the  precursor system can influence the  composition of the
nova ejecta.  It  is clear now that all the  possibilities for an {\sl
in situ}  formation of  the abundance pattern  found in the  ejecta of
novae must be rejected for  several reasons.  First, the amount of CNO
products observed in  the ejecta of novae is much  greater than can be
expected to characterize the matter  accreted from a companion star of
solar-like  composition.   Second,  breakout  from the  CNO  cycle  to
produce nuclei  like Ne, Na, and  Mg cannot occur  at the temperatures
smaller  than  300  million  degrees achieved  in  nova  thermonuclear
runaways. Third, the  abundance pattern of the ejecta  of novae cannot
be explained  by a  previous case BB  mass-transfer episode.   It then
follows that the explanation of the enriched heavy element composition
of the ejecta of classical novae must involve some degree of dredge-up
of matter from the underlying CO or ONe white dwarf.

Helium  and carbon-burning  reactions that  occur in  the core  of the
primary  during the  prior  evolutionary history  of  the white  dwarf
progenitor  can  explain the  formation  of  the  neon and  the  heavy
elements that  characterize the abundance  pattern of the  ejecta from
ONe novae.   Thus the importance of reliable  composition profiles for
massive   white   dwarfs,   resulting   from   detailed   evolutionary
calculations,  becomes clear.   In this  context, it  is  worth noting
that, due  to the lower  carbon abundance in  the outer layers  of ONe
white dwarfs, these objects are able to accrete more massive envelopes
(Jos\'{e}  \&   Hernanz,  1998)  between   successive  outbursts  and,
therefore,  the degree  of degeneracy  reached  at the  bottom of  the
accreted layers is higher and  the explosions are more violent than in
CO white dwarfs  of similar masses.  The higher  temperatures that ONe
white dwarfs in novae are able to reach, together with the presence of
``seed''  nuclei with  $A \geq  20$,  have also  consequences for  the
nature of  the material processed during  the outbursts: specifically,
we expect  that heavier  nuclei can be  formed in  these environments.
However, due  to the  relatively small masses  of nova ejecta,  and in
spite of  their moderately  high frequency, novae  do not  represent a
significant  source of  heavy element  enrichment of  Galactic matter.
Finally it  should be said  that although the  characteristic isotopes
ejected  by  novae involving  CO  and  ONe  white dwarfs  are  similar
($^{13}$C, $^{15}$N,  and $^{17}$O), their  ejecta have distinguishing
imprints: e.g. $^7$Li is produced in greater amounts by CO novae while
both  $^{22}$Na and  $^{26}$Al are  understood to  be produced  by ONe
novae (Jos\'e \& Hernanz 1998).

The mechanism responsible for  the outward mixing (dredge-up) of white
dwarf  core  matter into  the  hydrogen-rich  envelope  remains to  be
clearly  identified.  A  variety  of  possible  mechanisms  have  been
examined,  including  shear-driven  turbulent  mixing  (Kippenhahn  \&
Thomas 1978), Eckman spin-up (Livio \& Truran 1987), diffusion induced
core  convection  (Prialnik  \&  Kovetz  1984),  dredge-up  driven  by
convective overshoot (Glasner et  al.  1997), and most recently mixing
by wind-driven gravity wave breaking on the surface of the white dwarf
(Rosner et al. 2001; Calder et  al. 2002). Although it is true that it
is not  yet known how  much material is  mixed, it is also  clear that
reliable initial models  for the masses and compositions  of the white
dwarf components  of nova systems,  as a function  of the mass  of the
progenitor,  are required  for a  full understanding  of  the chemical
composition of the ejecta from novae.

In this paper we recompute the frequency of {\sl white dwarfs with ONe
cores} in novae, taking into account both the observational  selection
effects and our recent evolutionary  sequences for binary systems with
intermediate-to-massive  primaries (Gil-Pons \& Garc\'\i a-Berro 2001;
2002).  As will be shown  below, we expect  that  some  novae  hosting
massive  white  dwarfs  (e.g., an ONe  degenerate  core)  will show Ne
enrichments while others will not, due to the presence of a relatively
thick  CO  buffer.  We  also   discuss  the   consequences   of  these
evolutionary  sequences,  and in particular of this CO buffer, for the
expected  abundance  pattern of the ejecta.  The paper is organized as
follows.  In section 2, we  briefly  comment  on the  standard  double
common  envelope  scenarios  leading to novae that host massive  white
dwarfs.  Section 3 presents  our main  results for the remnant  masses
and their  compositions.  In section 4, we discuss  the  observational
selection effects arising from both the recurrence  intervals  between
successive  nova  outbursts  in  individual  systems  and the  spatial
distribution  of novae in the Galaxy.  In section 5, we comment on the
expected  abundances of the nova ejecta.  Discussion  and  conclusions
are presented in section 6.


\section{Scenarios leading to novae}

Evolutionary  sequences   for  primaries  of   mass  $10.0\,  M_\odot$
(Gil--Pons \& Garc\'\i a--Berro,  2001) and $9.0\, M_\odot$ (Gil--Pons
\& Garc\'\i  a--Berro, 2002) have  previously been computed.   For the
present work,  we have computed additional  evolutionary sequences for
primaries  of  masses  $M_{\rm  ZAMS}=10.5$, 11.0,  11.5,  and  $12.0\
M_\odot$.   The  physical inputs  are  the  same  as in  Gil--Pons  \&
Garc\'\i a--Berro  (2001) and  the results are  very similar  to those
presented there for  the whole range of masses.   The only significant
difference with respect to our previous work occurs for the primary of
$12.0\,  M_\odot$,  which ignites  carbon  at  its  center (see  \S  3
below). For binary systems hosting  primaries with a mass smaller than
$9.0\, M_\odot$ we have adopted  the results of Iben \& Tutukov (1985)
and  Iben (1990).  Our  systems  undergo the  so-called  case BB  mass
transfer (Delgado \& Thomas, 1981; de Loore \& Doom, 1992).  The first
Roche  lobe overflow  (RLOF) from  the primary  component  occurs when
hydrogen burning has  begun in a shell and  a deep convective envelope
forms around the helium core.  Mass loss in the presence of convection
induces unstable mass transfer between the components and leads to the
formation of  a common envelope  (Paczy\'nski, 1976).  This,  in turn,
induces   orbital  shrinkage.    The   primary  loses   most  of   its
hydrogen-rich envelope during the first  RLOF, and most of the mass is
lost from the system.  The  second RLOF episode leaves a remnant which
consists of  an ONe core surrounded by  a CO buffer and  a thin helium
layer.

\begin{figure}[t]
\vspace{7.5cm}
\hspace{-3.9cm}    
\includegraphics{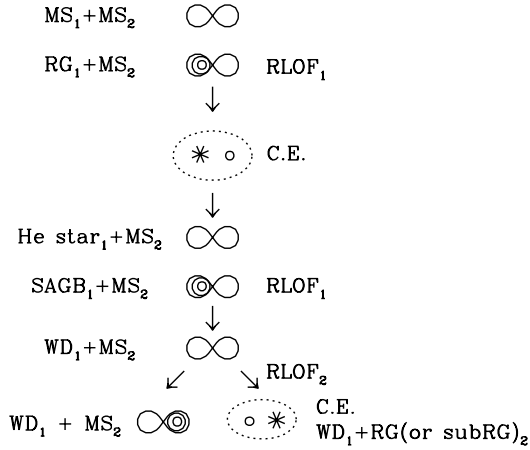}
\caption{Outline of pre--nova  scenarios hosting massive white dwarfs.
	{\sl MS,  SAGB}  and {\sl WD}  represent,  respectively,  main
	sequence, super--asymptotic giant branch star and white dwarf.
	{\sl RLOF} and {\sl C.E.}  refer to Roche  lobe  overflow  and
	common envelope phase, and the subindices  {\sl 1} and {\sl 2}
	correspond  to  the  primary  and  the  secondary   component,
	respectively.}
\end{figure}

Roche  lobe  overflow  from  the  secondary  (and  thus  reverse  mass
transfer) can occur, whether due  to further evolution of this star or
to angular  momentum losses. A nova  can only occur  for suitable mass
transfer  rates, specifically  $\lesssim  10^{-8} M_\odot\,$yr$^{-1}$,
However, systems in  which the mass transfer rates  are larger are not
necessarily prevented from evolving to novae.  In particular, Politano
(2002) has suggested  the possibility that such  systems might undergo
first a violent phase during  which the mass transfer rates are large,
ensuing important  mass and angular  momentum losses from  the system,
and then a  second phase, involving mass transfer  rates low enough to
allow the development of  nova outbursts.  Another possibility for the
occurrence of nova  outbursts involving a phase of  high mass transfer
rates  has been  discussed  by  Shara \&  Prialnik  (1994), and  Shara
(1994), who suggested the build-up of an ONe-rich layer on top of a CO
white dwarf as  a consequence of weak hydrogen  and helium flashes.  A
subsequent phase of low mass accretion rates would then trigger a nova
explosion.  However,  fine tuning of model parameters  is necessary to
obtain a  realistic nova event,  for the conditions proposed  by these
authors.  It thus seems  that, while  we may  have a  good qualitative
picture of the nature and  evolution of the progenitors of novae, many
detailed  features of  their  evolution remain  to  be fully  explored
numerically. A schematic  view of our scenario is  presented in Figure
1.


\section{The degenerate cores}

\begin{figure}[t]
\vspace{8.0cm}
\hspace{-2.5cm}    
\includegraphics{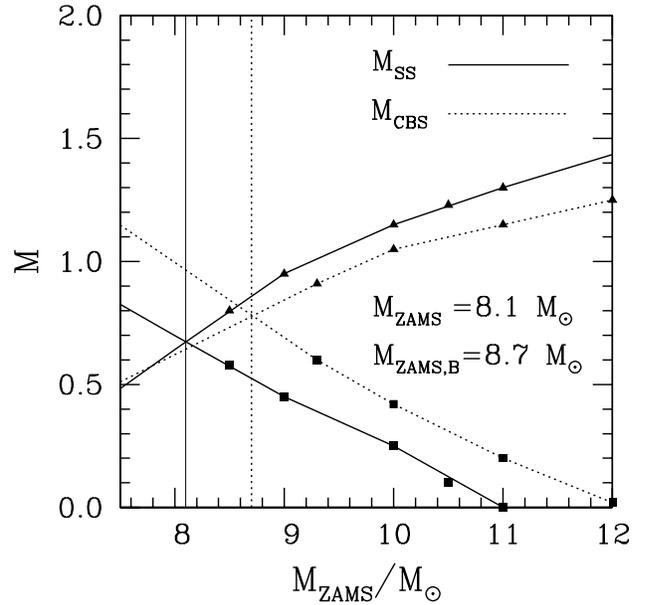}
\caption{Size  of the CO cores  at the  beginning  of  carbon  burning
	 (triangles)  and  mass  point  at  which  carbon  is  ignited
	 (squares).  The results for the  evolution  of a single  star
	 are represented as solid lines, whereas dotted lines show the
	 results for the evolution in a CBS.  See text for  additional
	 details.}
\end{figure}

The size  of the CO core just  before carbon burning sets  in, and the
mass  at which  carbon ignition  occurs are  shown in  Figure 2,  as a
function of the  mass on the zero age main  sequence (ZAMS); these are
shown both for the case of single stars (solid lines) and for stars in
close binary  systems (dotted lines).   These cores are the  result of
considering a case BB  non--conservative mass transfer episode between
the  components.  The first  mass  transfer  episode  occurs when  the
primary star is ascending the  red giant branch, and a deep convective
envelope is  formed. In  such a scenario,  a common envelope  phase is
very  likely to  occur, and  significant amounts  of mass  and angular
momentum can  be lost  from the system.  Systems initially  closer may
undergo  mass  transfer while  the  donor  is  still surrounded  by  a
radiative  envelope.  In  this  case, we  would  obtain somewhat  more
massive  ONe white  dwarfs.   The  differences we  have  found in  our
calculations  are  about  $0.02  \,  M_\odot$  between  the  ONe  core
resulting  from the conservative  and non--conservative  mass transfer
case,  whereas   the  differences  between   considering  single  star
evolution and evolution in a  close binary are about $0.1 \, M_\odot$.
Hence, for  reasonable choices of  the orbital parameters the  size of
the  resulting  He core  is  quite  insensitive  to the  computational
details (Gil--Pons \& Garc\'\i  a--Berro, 2002). It follows that, even
though the initial-to-final mass relation of the primary components is
not unique  but depends to some  extent on the  initial separation and
mass ratio  of the binary, these differences  are considerably smaller
than  those resulting  from taking  into account  the  consequences of
binary evolution.

\begin{figure}[t]
\vspace{8.0cm}
\hspace{-2.5cm}    
\includegraphics{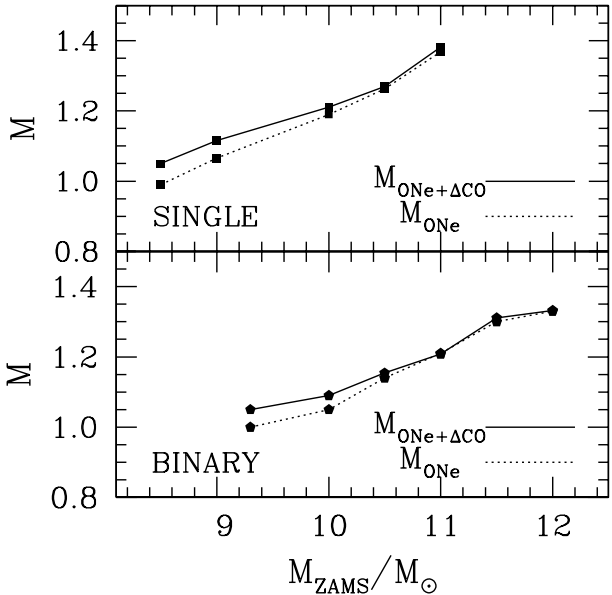}
\caption{Size  of the final  cores as a function  of the ZAMS mass for
	 single and binary star evolution.}
\end{figure}

As one can  see, the remnant cores of close  binary components tend to
be  smaller.  This  behavior  has  also been  found  in several  other
theoretical works  --- see, for  instance, Iben \& Tutukov  (1985) and
references therein.   Note that  for this range  of masses,  carbon is
ignited off-center and that, as the mass of the primary increases, the
mass at which carbon is ignited approaches the center.  In particular,
we find that for the case of isolated evolution, stars of mass $M_{\rm
ZAMS}\,\simeq  \, 11\,M_\odot$  ignite carbon  at the  center, whereas
this limit turns  out to be $M_{\rm ZAMS}\,\simeq  \, 12\,M_\odot$ for
the case of the evolution in a CBS.  We also note that, as the mass of
the star decreases from $12 \, M_\odot$ to $9 \, M_\odot$, the mass of
the degenerate CO core correspondingly decreases, whereas the point at
which  carbon  is  ignited  approaches the  He-C  discontinuity.   The
intersection of  the two curves  determines the minimum  mass required
for carbon  ignition to  take place.  The  lines intersect  at $M_{\rm
ZAMS}\,  \simeq\,  8.1\,  M_\odot$ and  $8.7\,M_\odot$,  respectively.
This represents  the lower mass limit  for the formation  of ONe white
dwarfs.   However, this  mass cannot  be  taken as  the limiting  mass
separating  CO from  ONe  white  dwarfs, since  the  fact that  carbon
ignition occurs does not necessarily imply that burning is going to be
extended enough in  the CO core to change  drastically its composition
and form an ONe white dwarf.  In fact we find that the limiting masses
are  somewhat larger,  $9.0\,M_\odot$ and  $9.3\,M_\odot$ respectively
and that  for masses smaller  than these the carbon  burning reactions
never proceed at a significant  rate.  Note, however, that in any case
we always obtain smaller cores for the evolution in CBSs.  This result
influences  the expected  frequency  of ONe  white  dwarfs, since  the
larger value of  the minimum mass allowing for  the formation of novae
hosting an ONe white dwarf  obtained here, combined with the fact that
IMF tends to favour the  formation of low-mass stars, will also favour
the  formation of  CO novae.   The effects  of binary  evolution, even
though they  may be weaker  than the observational  selection effects,
will thus act against novae hosting an ONe white dwarf.

\begin{table}[t]
\caption{Size of  the final cores as  a function of the  ZAMS mass for
         binary star evolution.}  
\centering
\begin{tabular}{lrc}
\hline 
\hline 
$M_{\rm ZAMS}$  & 
$M_{\rm ONe}$ & 
$M_{{\rm ONe}+ \Delta{\rm CO}}$ \\ 
\hline  
 9.3 & 1.00 & 1.07\\ 
10.0 & 1.05 & 1.09\\ 
10.5 & 1.14 & 1.15\\ 
11.0 & 1.21 & 1.22\\  
11.5 & 1.30 & 1.31\\ 
12.0 & 1.33 & 1.33\\ 
\hline 
\hline
\end{tabular}
\end{table}

Another factor which influences the expected frequency of nova systems
hosting an ONe white dwarf is the relationship between the initial and
final mass.  The degenerate cores  obtained in our simulations have an
ONe core surrounded  by a CO buffer.  Figure  3 shows the relationship
between the  initial mass  of the  primary and the  mass of  the white
dwarf (solid  line), and the  size of the  ONe core (dotted  line) for
both the single  and the binary cases.  For  the sake of completeness,
these two quantities are also presented in Table 1.

With respect  to the final  chemical composition profiles  obtained in
our  evolutionary sequences  after carbon  has been  exhausted  in the
central regions,  the most important characteristic  of the degenerate
cores is the  existence of two different regions:  an inner core where
oxygen  and neon are  the dominant  constituents, and  a carbon-oxygen
layer  surrounding  this  core,  where extensive  helium  burning  has
occurred, but  where the physical  conditions are not  compatible with
carbon burning.  As one can see in  Table 1, the size of this CO layer
decreases as the total white dwarf mass increases, from $\Delta M_{\rm
CO}=0.07\, M_{\odot}$  for the remnant  of the $9.3 \,  M_\odot$ star,
down to $\Delta  M_{\rm CO} \,< \,0.01 \,M_\odot$,  for the remnant of
the $12 \,  M_\odot$ primary.  The presence of this  CO buffer has two
foreseeable  consequences  for the  nova  outbursts.   First, one  can
expect that the first nova outbursts occurring on such a core will not
show substantial  neon enrichments.  Depending upon  the efficiency of
the mixing between the material of  the inner core and of the accreted
material, neon might not be detectable  until most of the CO buffer is
dredged-up and ejected.  Also, the  presence of an important amount of
carbon increases  the rate of energy  generation by the  CNO cycles at
lower  temperatures   and  thus  ensures  an   earlier  occurrence  of
thermonuclear runaway.  A ``naked'' ONe  white dwarf would be  able to
accrete more mass  between outbursts and would allow  for more massive
ejecta (Jos\'e et al.  2002).


\section{The observed rates of oxygen--neon nova outbursts}

The observed fraction of nova outbursts in which an ONe white dwarf is
involved depends  upon the relative  frequency of ONe white  dwarfs in
close  binary systems  and upon  several selection  effects.   We will
consider here selection effects of two types.  First we will take into
account those related to the recurrence time between outbursts.  Since
massive white dwarfs  --- and, hence, ONe white  dwarfs --- experience
nova outbursts more  frequently, there will be a  clear selection bias
favouring their detection.  Second,  more massive white dwarfs produce
more luminous outbursts, thus favouring their detection as well.

In  deriving the frequency  of nova  outbursts in  which an  ONe white
dwarf is involved, we follow closely the treatments of Truran \& Livio
(1986) and  Ritter et  al.   (1991).   We assume  that  a white  dwarf
undergoes  a thermonuclear  flash  when a  certain critical  pressure,
$P_{\rm crit}$, is reached.  A  good approximation to this quantity is
(MacDonald, 1983):

\begin{figure}[t]
\vspace{8.0cm}
\hspace{-2.5cm}    
\includegraphics{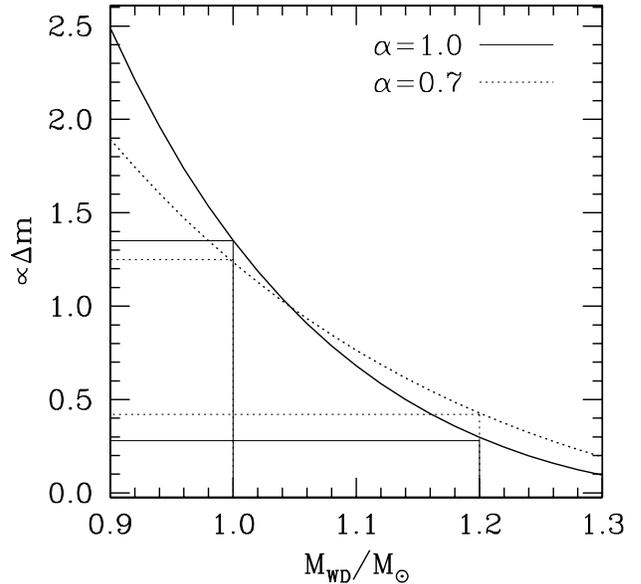}
\caption{Mass accreted  between outbursts as  a function of  the white
         dwarf mass}
\end{figure}

\begin{equation}
P_{\rm crit}=\frac{G M_{\rm WD} \Delta m}{4 \pi R_{\rm WD}^4}
\end{equation}

\noindent where $M_{\rm WD}$ and $R_{\rm WD}$ are the white dwarf mass
and radius, respectively,  and $\Delta m$ is the  critical mass of the
accreted layer (Truran  \& Livio, 1986).  The amount  of accreted mass
necessary for the ignition to take place can be obtained from a fit to
full numerical simulations (Fujimoto, 1982; Prialnik et al., 1982):

\begin{equation}
\Delta  m  \propto   
\Bigg(\frac  {M_{\rm  WD}}{{R_{\rm  WD}}^4}\Bigg)
^{-\alpha}{\dot M_{\rm WD}}^{-{\beta}}
\end{equation}

\noindent where  $\alpha$ and $\beta$  are parameters to be  fitted to
detailed  nova simulations,  and $\dot  M_{\rm WD}$  is  the accretion
rate.  The calculations of Fujimoto  (1982) and Prialnik et al. (1982)
show only a weak dependence on the mass accretion rate (Ritter et al.,
1991) and, therefore, we will assume  $\beta = 0$.  This is tantamount
to  assuming that  the recurrence  time  does not  influence the  nova
distribution (Ritter  et al.  1991).   Note as well  that calculations
(MacDonald 1983),  show a weak  dependence on the  initial luminosity,
which will  also be  neglected here.  On  the other hand,  $\alpha$ is
typically $\simeq  1.0$ (Truran  \& Livio, 1986).   However, numerical
calculations which  include a detailed treatment  of diffusion (Kovetz
\& Prialnik,  1985) show that  $\alpha=0.7$ may be a  more appropriate
value.   Therefore, we will  perform calculations  for both  $\alpha =
1.0$ and $\alpha =0.7$.

Fig.  4  shows the  masses of the  layers accreted  between successive
outbursts  as a  function of  the mass  of the  white dwarf,  for both
choices of $\alpha$.  The thin  lines indicate the accreted masses for
a  $1.0\, M_\odot$  and a  $1.2 \,  M_\odot$ white  dwarf,  which are,
respectively, the  lower mass limit  allowing for the formation  of an
ONe white  dwarf and a representative  value of the mass  of ONe white
dwarfs.  As one can see, the  mass of the accreted layers changes by a
factor  of roughly 4.5  for this  range of  masses if  $\alpha=1.0$ is
used, whereas this change is slightly  smaller (only a factor of 3) if
$\alpha=0.7$ is  adopted.  If we  assume that mixing is  not efficient
enough to  allow for  significant neon enrichments  until most  of the
CO--rich layer is removed, the  resulting mass of the white dwarf will
be smaller when  the first neon enrichments show  up.  Thus, more mass
needs to be accreted in order  to produce an outburst, as clearly seen
in Fig. 4. This could possibly lead to a correlation between high neon
and heavy  metal enrichments and  high ejected masses, as  observed in
some novae, but  detailed calculations, which are out  of the scope of
this paper,  should be carried out  in order to check  whether this is
true or not.

The  time interval  between outbursts  can  be computed  as the  ratio
between the mass accretion rate and  the mass of the layer required to
reach $P_{\rm  crit}$.  Following Ritter  et al.  (1991),  a selection
function is defined as:

\begin{equation}
S_{\rm N}(M_{\rm WD}) = 
L^n(M_{\rm WD})\Bigg(\frac {M_{\rm WD}}{R_{\rm WD}^4}\Bigg)^\alpha
\end{equation}

Here $L(M_{\rm WD})$  is the luminosity of the  hydrogen burning shell
--- for which we  adopt the prescription of Kippenhahn  (1981) --- and
$R(M_{\rm WD})$  is determined using the  mass--radius relationship of
Nauenberg  (1972).   The exponent  $n$  in  Eq.   3 accounts  for  the
possible  spatial   distribution  of  novae  (Ritter   1986).   For  a
volume-limited sample  the exponent is  $n = 0$;  the value $n  = 1/2$
corresponds  to a  flux-limited sample  from a  disk-like distribution
where interstellar extinction is taken into account; the value $n=3/4$
corresponds to  a flux-limited  sample from an  isotropic distribution
where  interstellar extinction is  taken into  account (Ritter  et al.
1991).

\begin{table}
\caption{Frequency of occurrence of novae of different types.}
\centering
\begin{tabular}{llccccc}
\hline
\hline
{}&{}&\multicolumn{2}{c}{SS} &&\multicolumn{2}{c}{CBS}\\
\cline{3-4}
\cline{6-7}
    {}
  & {}
  & {$P_{\rm CO}$}
  & {$P_{\rm ONe}$}
  & {}
  & {$P_{\rm CO}$}
  & {$P_{\rm ONe}$}\\
\hline
$\alpha = 1.0$ & $n$ = 0     & 0.54 & 0.46 && 0.68 & 0.32\\
               & $n$ = 1/2   & 0.51 & 0.49 && 0.63 & 0.37\\
               & $n$ = 3/4   & 0.49 & 0.51 && 0.58 & 0.42\\
$\alpha = 0.7$ & $n$ = 0     & 0.55 & 0.45 && 0.67 & 0.33\\
               & $n$ = 1/2   & 0.51 & 0.49 && 0.63 & 0.37\\
               & $n$ = 3/4   & 0.50 & 0.50 && 0.60 & 0.40\\
\hline
\hline
\end{tabular}
\end{table}

The relative  probability of  observing an outburst  from a nova  on a
white dwarf which descended from  a main sequence star of mass $M_{\rm
ZAMS}$ is given by

\begin{equation} 
P(M_{\rm WD}) = S_{\rm N}(M_{\rm WD}) \times {M_{\rm ZAMS}}^{-2.35}
\end{equation}

\noindent  where we have assumed a Salpeter IMF  (Salpeter,  1955).  A
more precise  treatment of the problem  should  include a  statistical
study of the  orbital  parameters  of the  binaries  leading  to novae
(Politano, 1996) and of the effects of mass loss from the white dwarf,
but for our purposes the approach  adopted here is adequate.  With all
these inputs our results for the frequencies of occurrence of novae of
the  different  types are shown in Table 2.  Columns  3 and 4 give the
occurrence  frequencies  for the case for which  the  initial-to-final
mass relation obtained for the evolution of isolated stars is adopted;
columns  5 and 6 show the  results  obtained  for  evolution  in close
binaries.  Note that when the results for the  evolution  of  isolated
stars are adopted, the  frequency of novae  hosting an ONe white dwarf
is always larger than that obtained for the evolution of a CBS.  Thus,
the  effects  associated  with  binary  evolution  act to  reduce  the
fraction of ONe white dwarfs in nova systems relative to a single star
distribution.  This stems  directly  from the fact that the  masses of
the cores are considerably smaller for the evolution in a CBS.

With regard  to the  effects of the  spatial distribution it  is worth
noticing that: (1)  for the case of a  volume-limited distribution the
fraction of novae  in which an ONe white dwarf  is involved is roughly
1/3;  (2)  for a  disk-like  distribution  this  fraction is  slightly
higher; and (3) for the case of an isotropic distribution it increases
up  to a value  of 40\%.  This means  that the  choice of  a disk-like
distribution,   which  is   more  appropriate   to  account   for  the
distribution  of Galactic  novae,  also increases  the selection  bias
favouring more massive  white dwarfs.  This trend is  even clearer for
an isotropic  distribution in  which interstellar absorption  is taken
into account  ($n=3/4$).  Finally,  the effects of  adopting different
values  for $\alpha$  (that  is the  effects  associated to  different
choices  of  the  critical   pressure)  are  much  weaker  than  those
associated with considering different spatial distributions (different
values of  $n$).  


\begin{figure}[t]
\vspace{8.0cm}
\hspace{-2.5cm}    
\includegraphics{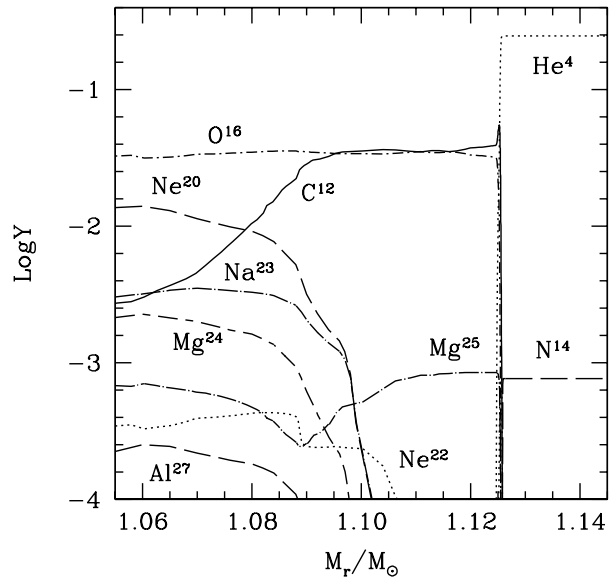}
\caption{Number  abundances of the  uppermost  regions of an ONe white
	 dwarf  resulting from the evolution of a $10 M_{\odot}$  ZAMS
	 primary component in a close binary system.}
\end{figure}

\section{The chemical composition of the ejecta}

Fig.  5 shows  the profiles of the chemical  abundances of the remnant
of  the $10\,  M_ \odot$  primary component  of a  CBS.   As discussed
previously, this profile  is characterized by an ONe  rich core on top
of which there  is a CO rich buffer.  The existence  of this layer has
consequences for  the expected abundances  of the nova ejecta,  as the
innermost material cannot be dredged-up and expelled until significant
erosion of the surrounding CO  layer has occurred, due to earlier nova
outbursts.  It is not easy to give a precise estimate of the number of
outbursts that must  occur before a significant amount  of neon can be
observed, since  there are two  factors which are not  well understood
and play an important role.  The first factor is the mixing efficiency
in  the uppermost layers  of the  white dwarf:  how fast  this process
operates and  how much material  synthesized during carbon  burning is
dredged up  to its  surface.  The second  factor, of a  very different
nature,  has  to  do  with  the  possibility of  a  variation  in  the
recurrence time.  Even though it  is often assumed that the recurrence
time  remains constant,  there  are  reasons to  suspect  that it  may
change.

First of  all, changes  in the mass  transfer rate from  the secondary
that may occur  as a consequence of the nova  outbursts can modify the
recurrence time.  However,  Kolb et al.  (2001) have  shown that these
nova--induced mass  transfer variations do not play  an important role
if mass transfer is sufficiently stable.  There is another possibility
for  changes in  the recurrence  time, that  has to  do  with eventual
changes  of the  white  dwarf mass.   These  changes are  not easy  to
quantify  but the  fact that  some nova  abundance patterns  show high
metal enhancements, together with our  results for the ONe white dwarf
composition profiles (ONe core  surrounded by CO buffer), suggest that
erosion  of the  surface  material of  the  degenerate component  must
exist,  and  therefore,  the  white  dwarf mass  must  decrease.   The
chemical  composition of the  surrounding layers  can also  modify the
recurrence  time in  the  following  sense: if  the  amount of  carbon
increases,  the  reactions of  the  CNO  cycle  are enhanced  and,  as
demonstrated by Jos\'{e} \& Hernanz (1998), a smaller amount of matter
needs  to  be  accreted  between  outbursts  in  order  to  develop  a
thermonuclear runaway.  Hence, the recurrence time decreases.

\subsection{The carbon--oxygen rich layer}

For comparison purposes, we present in Fig. 6 the profiles of chemical
abundances at the outer shells of a CO white dwarf.  The main isotopes
are carbon and oxygen, as they were for the surface (buffer) region of
the ONe white dwarf, but the different prior evolutionary histories of
these remnants have  left their imprints in the  compositions of other
isotopes.   In  particular,  non-negligible abundances  of  $^{20}$Ne,
$^{23}$Na and $^{24}$Mg  appear at the innermost part  of the CO layer
of an ONe white dwarf, whereas these elements are absent in a CO white
dwarf.  Such  composition  differences  should  be  kept  in  mind  in
interpretations  of the  observed abundance  patterns in  nova nebular
remnants.

It is also  interesting to compare the CO-rich  buffers of the remnant
of  the  $10\,  M_\odot$ ZAMS  primary  (Fig.   5)  with that  of  the
corresponding single star --- Fig.  9 in Ritossa et al.  (1996) --- in
order to illustrate the importance of our consideration of the effects
of binary evolution. The most significant difference is in the size of
the CO buffer layer, that may be seen to be thicker for the remnant of
the primary in a CBS ($0.07\, M_\odot$ versus $0.01\, M_\odot$ for the
isolated  white dwarf). Since  the relative  size of  the ONe  core is
important to considerations of  the relative frequencies of occurrence
of  CO versus  ONe novae,  the  change of  about 0.1  M$_\odot$ is  of
interest. Moreover, the remnant of the binary component has a somewhat
larger carbon abundance than that of the single ONe white dwarf, which
can  influence the  runaway timescale  and  the size  of the  accreted
envelope.

\begin{figure}[t]
\vspace{8.0cm}
\hspace{-2.5cm}
\includegraphics{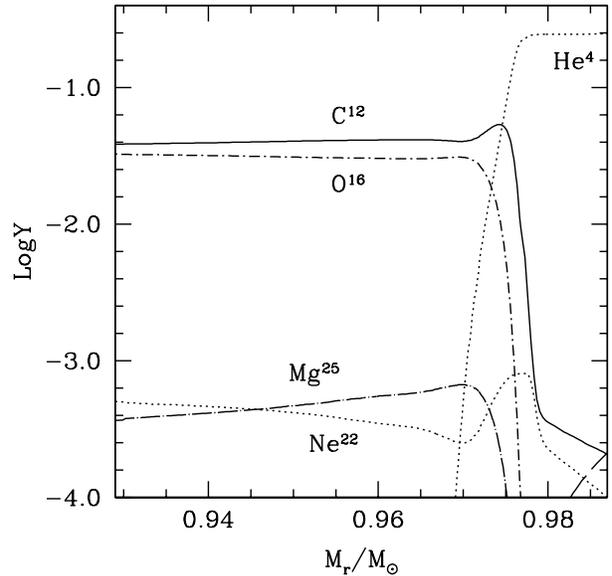}
\caption{Number  abundances of  the CO  core  resulting from  a $9  \,
         M_\odot$ ZAMS primary component in a CBS}
\end{figure}

\subsection{The oxygen-neon cores}

The comparison of the inner ONe cores obtained from the evolution of a
$10 \,  M_\odot$ single  star with the  remnant of the  $10\, M_\odot$
primary component  of a CBS shows  that the final  masses are slightly
different ($1.18\, M_\odot$ and $1.06\, M_\odot$, respectively).  This
implies that neon--enriched outbursts  will occur in a less degenerate
medium  than  that  expected  if  single evolution  is  followed  and,
therefore, they will be less  violent.  In fact we obtain a degeneracy
parameter at the bottom of the CO buffer which is considerably smaller
($\sim 33\%$)  for the case  of the remnant  of the primary star  of a
CBS. Additionally,  the critical mass  $\Delta m$ needed to  develop a
thermonuclear runaway also changes.  This effect can be estimated with
the help  of Fig.  4. As  one can see,  the critical mass that  can be
accreted by a  1.06 $M_\odot$ white dwarf (the remnant  of the CBS) is
about 2.6 times larger than the  critical mass that can be accreted by
a  1.18  $M_\odot$ white  dwarf  (the  remnant  of the  single  star).
Consequently, the possibility of accreting more mass between outbursts
arises.   This may  help us  to understand  an  apparent observational
trend that  is somewhat of a  challenge to theory: the  fact that high
neon  and  heavy  element   enrichments  are  sometimes  found  to  be
associated  with novae with  larger ejecta  masses. Subsequent  to the
dredge-up and  ejection of  the overlying carbon-oxygen  buffer layer,
the  masses of the  residual ONe  cores we  have calculated  allow the
possibility  of accreted  envelopes  approaching 10$^{-4}\,  M_\odot$.
Moreover, if  the remaining parameters of interest  are kept constant,
the recurrence time  between outbursts will change as  the mass of the
critical  accreted layer and,  therefore, the  frequency in  this case
will decrease by roughly a factor of $2.0-2.6$.


\section{Discussion and conclusions}

The calculations described in this paper may be summarized as follows:

\begin{itemize}

\item We have computed  self-consistent evolutionary sequences for the
formation  of white  dwarfs with  ONe  cores in  close binary  systems
involving  intermediate mass stars.   We have  found that  the minimum
mass at the  ZAMS which allows for extensive carbon  burning in the CO
core is $\simeq \, 9.3 \, M_\odot$, and that the mass of the resulting
white dwarf  is, in this case,  $M_{\rm WD} \simeq  \, 1.1\, M_\odot$.
This value is somewhat lower than previous limits on the masses of ONe
white dwarfs.

\item We have obtained the initial-to-final mass relationship and used
our  evolutionary sequences to  revisit the  question of  the relative
fraction  of novae  which contain  white dwarfs  with ONe  cores.  Our
motivation to reconsider this problem  has been the major influence of
the  white  dwarf  mass  (through  the mass-radius  relation)  on  the
selection effects that favour the observation of outbursts produced in
more massive novae.  In our determination of the relative frequency of
novae  with ONe  cores, we  have  followed closely  the treatments  of
Truran \& Livio (1986) and Ritter et al.  (1991).

\item We have found that the  primary effect of binarity consists of a
noticeable  reduction in  the  frequency of  novae  hosting ONe  white
dwarfs.   Specifically, when  we  take into  account  our results  for
evolution in a CBS, we find that approximately $37 \%$ of the observed
nova outbursts are  expected to host ONe white  dwarfs for a disk-like
distribution.  This fraction decreases to $32 \%$ for a volume-limited
sample.  Since  the Galactic  distribution of classical  novae extends
beyond the thickness of  the Galactic disk and interstellar absorption
plays an important role,  the relevant distribution is therefore given
by a disk-like  distribution --- that is, $n=1/2$  in Eq.  (3).  These
results  imply that flux  limitation favours  observation of  the most
luminous novae: those occurring on the most massive white dwarfs.

\item We have also  calculated detailed composition profiles for these
remnants. This  is one of  our major results.  We have found  that the
remnants of  the evolution  in CBS always  have a relatively  thick CO
buffer on top of the ONe  core.  This means that novae hosting massive
white dwarfs with ONe cores can show Ne enrichments only if either the
CO buffer has been removed  by successive nova outbursts or been mixed
with material from the ONe core by some other mechanism.  In contrast,
if  the   mass  transfer  process  has  started   only  recently,  the
thermonuclear runaway  will proceed  on top of  the CO buffer  and the
ejecta will  reflect the  abundance pattern of  a massive CO  nova. Of
course, ultimately  the CO  surface layer will  be eroded and  the ONe
core will be exposed.

\item As pointed out by  Livio \& Truran (1994), the interpretation of
the  observed yields  of  novae remains  somewhat uncertain.   Stellar
evolution theory clearly predicts that, at birth, ONe white dwarfs are
more  massive  than CO  white  dwarfs.   However,  the rather  massive
nebular remnants  inferred for some novae with  large neon enrichments
are difficult  to reproduce on  the basis of current  theoretical nova
simulations.  Our models  seem to give some hints  towards a (partial)
solution to this  problem.  First, we have pointed  out that ONe white
dwarfs  do not  have a  homogeneous composition  profile  but, rather,
consist of  two clearly different  regions.  The dominant  elements in
the  inner zone  are oxygen  and neon,  while the  overlying  layer is
mainly composed of  carbon and oxygen. This CO buffer  turns out to be
important when interpreting  the nature of the nova  ejecta, since its
size is not the same for all  ONe white dwarfs: it is smaller for more
massive white  dwarfs (see Table 1).   When the CO  buffer is expelled
after a series of outbursts, we are left with a naked ONe white dwarf,
whose typical mass is between  approximately 1.0 and 1.1 $M_\odot$ but
ranges up to the Chandrasekhar  limit.  For the lowest mass cases, the
white  dwarf cores can  even be  somewhat less  massive than  the most
massive CO  white dwarfs, and thus  can be expected  to accrete rather
massive envelopes  before the thermonuclear runaway  ensues.  This may
help to explain the rather large  masses that have been quoted for the
ejecta of such identified ``neon novae''  as V1370 Aql 1982 and QU Vul
1984 (Saizar et al., 1992; Andre\"a et al., 1994; Schwarz, 2002).

\end{itemize}

It seems clear that the last  word on these questions has not yet been
told, but we  believe that our results for  ONe white dwarfs resulting
from close binary  evolution will help to provide a  solid basis for a
better understanding of nova explosions.

\begin{acknowledgements} 
Part of  this work  was supported by  the Spanish DGES  project number
PB98--1183--C03--02/03,    by   the    MCYT    grants   AYA2000--1785,
AYA2001--2360,  and  by the  CIRIT.   JWT  would  like to  acknowledge
support  by the  US  Department  of Energy  under  Grants B523820  and
DE-FG02-91ER40606 at the University of  Chicago. We also would like to
acknowledge our referee, H. Ritter, for his very valuable suggestions.
\end{acknowledgements}



\end{document}